\documentclass[12pt]{iopart}
\usepackage[bookmarks,dvips,pdfhighlight=/O,pdfstartview=FitH]{hyperref}
\usepackage{iopams}
\usepackage{graphicx,setstack,amsopn,color}

\DeclareMathOperator{\im}{Im}

\newcommand{\dd}{\mathrm{d}}

\newcommand{\erw}[1]{\left \langle #1 \right \rangle}

\begin{document}

\title{The dilepton probe in heavy-ion collisions}

\author{Hendrik van Hees}

\address{Cyclotron Institute and Physics Department, Texas A\&M
  University, College Station, Texas 77843-3366, USA}

\ead{hees@comp.tamu.edu}

\begin{abstract}
  Dileptons provide direct observables of the electromagnetic
  current-current correlator in the hot and/or dense medium formed in
  heavy-ion collisions. In this article an overview is given about the
  status of the theoretical understanding of the dilepton phenomenology
  in heavy-ion collisions from studies of the in-medium properties of
  hadrons and partons within many-body theory and about connections to
  fundamental questions concerning the chiral phase transition.
\end{abstract}

\section{Introduction}

In heavy-ion collisions electromagnetic probes, i.e., photons and lepton
pairs (``virtual photons'') provide one of the most valuable
possibilities to study the interior of the hot and dense medium created
in the interaction over its whole history since their spectra are nearly
unaffected by final-state interactions~\cite{Rapp:1999ej}.

This paper will be restricted to invariant-mass ($M$) and
transverse-momentum ($q_T$) spectra of dileptons whose rate is given
by~\cite{Shuryak:1980tp,MT84,Gale:1990pn}
\begin{equation}
\label{MT}
\frac{\dd N_{ll}}{\dd^4 x \dd^4 q} = -\frac{\alpha^2}{3 \pi^3}
\frac{L(M)}{M^2} \im \Pi_{\mathrm{em}\mu}^{\mu}(M,q;T,\mu_B) f_B(q_0,T),
\end{equation}
where $\alpha\simeq 1/137$ denotes the fine-structure constant,
$M=q_0^2-q^2$ the invariant mass of the lepton pair of energy, $q_0$,
and three-momentum, $q$, $T$ the temperature, $\mu_B$ the baryon chemical
potential, $f_B$ the Bose distribution, and $L(M)$ the lepton-phase
space factor. As is known from $e^+ e^- \rightarrow \mathrm{hadrons}$,
in the vacuum the retarded hadronic electromagnetic (em.) current
correlator, $\Pi_{\mathrm{em}}$, at low invariant masses, $M \lesssim
M_{\mathrm{dual}}$ is well described by the vector-meson dominance (VMD)
model for the light vector mesons $\rho$, $\omega$, and $\phi$ and by
the perturbative QCD (pQCD) continuum at higher masses ($M \gtrsim
M_{\mathrm{dual}}$), where $M_{\mathrm{dual}} \simeq 1.5\,\mathrm{GeV}$
denotes a ``duality scale''. The hadronic ``resonance part'' is
dominated by the isovector channel ($\rho$ meson).

For a theoretical description of dilepton production in relativistic
heavy-ion collisions (HICs) thus the first goal must be an understanding
of the in-medium spectral properties of the light vector mesons in the
hadronic and the QGP in the partonic phase of the fireball
evolution. This review of recent progress in this field is organized as
follows: In Sect.~\ref{sect.emcc} first the constraints on the
em. current correlator in strongly interacting matter from QCD as the
underlying fundamental theory of strong interactions will be discussed,
followed by a brief summary on effective hadronic models. In
Sect.~\ref{sect.phen} these models are confronted with data from
ultrarelativistic HICs. Sect.~\ref{sect.concl} contains brief
conclusions and an outlook.

\section{The electromagnetic current correlator in strongly interacting matter}
\label{sect.emcc}

\textbf{Ab-initio constraints.} In the vacuum and at low temperatures
and densities the light-quark sector of QCD is governed by (approximate)
\emph{chiral symmetry} which is spontaneously broken by the formation of
a quarks condensate, $\erw{\bar{\psi} \psi} \neq 0$, in the QCD vacuum
which manifests itself in the mass splitting of chiral partners in the
hadron spectrum, as can be seen, e.g., in the experimental determination
of the isovector-vector and -axialvector current correlator through
$\tau \rightarrow \nu+n \pi$ decays~\cite{aleph98,opal99}. From
(lattice) QCD at finite temperature one expects a decrease of the quark
condensate at high temperatures and/or densities and restoration of
chiral symmetry. Thus the mass spectra of hadrons are expected to soften
and to degenerate with their pertinent chiral partners above a critical
temperature, $T_c$. One indication of the interrelation of chiral
symmetry and confinement is that in lattice-QCD (lQCD) calculations the
critical temperature, $T_c \simeq 160$-$190 \, \mathrm{MeV}$, for the
chiral and the deconfinement (cross-over) transitions
coincide~\cite{Karsch:1994hm}.

From hadronic modeling two microscopic mechanisms for chiral symmetry
restoration (CSR) have emerged: On the one hand it has been conjectured
that the hadron masses drop to zero at the critical point due to the
melting of the quark condensate)~\cite{Brown:1991kk}. On the other
hand within phenomenological hadronic many-body models the hadron spectra
show a significant broadening with little mass shifts
(``melting-resonance
scenario'')~\cite{Gale:1990pn,Gale:1993zj,Rapp:1997fs,Rapp:1999us}. A
direct relation between in-medium spectral properties of hadrons, in our
context particularly vector mesons, to QCD is provided by QCD sum rules
which relate moments of the (in-medium) spectral functions for various
currents in different isospin channels in the space-like region to the
pertinent quark and four-quark condensates. Detailed
studies~\cite{Asakawa:1993pq,Leupold:1997dg,Klingl:1997kf,Ruppert:2005id}
show that (in cold nuclear matter) both the ``dropping-mass'' and the
``melting-resonance'' scenarios for CSR are compatible with QCD sum
rules. One objective for the investigation of dileptons in high-energy
heavy-ion collisions is thus to gain insight in the in-medium spectral
properties of the light vector mesons through the em. current correlator
to constrain the mechanism leading to the softening of the spectral
functions. Due to the experimental problems to assess the spectral
properties of the axial-vector channel, an indirect theoretical approach
may be to connect chiral hadronic models, describing successfully the
measured dilepton observables, with CSR via finite-temperature
Weinberg-sum rules~\cite{Weinberg:1967kj,Kapusta:1993hq} which relate
moments of the \emph{difference of vector- and axial-vector-current
  correlators} to quark and four-quark condensates, i.e., order
parameters of chiral symmetry, providing constraints on these models
from lQCD.

\textbf{Hadronic many-body theory.} A (model-independent) approach to
assess medium modifications of vector (and axial-vector) mesons is based
on the chiral reduction formalism, providing a low-density expansion of
the in-medium vector- and axial-vector current correlators in terms of
the corresponding vacuum quantities which are taken from experimental
data such as $\tau \rightarrow n \pi \nu$. The most intriguing feature
of these kind of models is the ``in-medium mixing'' of the vector- with
the axial-vector current correlator due to pions in the
medium~\cite{Dey:1990ba} which may provide a mechanism for the onset of
CSR. However, the applicability of the low-density approximation is
restricted to very low temperatures and densities.

Another ansatz is to use chiral models in various realizations of chiral
symmetry. One possibility is to describe vector- (and axial-vector)
mesons as gauge bosons within the \emph{(generalized) hidden-local
  symmetry models}~\cite{bandokugo84,Harada:2003jx}. Here a particular
realization of chiral symmetry, the \emph{vector manifestation}, becomes
possible, where the chiral partner of the longitudinal $\rho$ meson is
the pion, thus providing a definite chiral model for the ``dropping-mass
scenario''. A detailed renormalization-group analysis shows that such a
model together with Wilson-matching of the effective model to QCD
(leading to an ``intrinsic'' temperature/density dependence of the
effective-model parameters), inevitably leads to a vanishing $\rho$ mass
at the critical point and a violation of VMD~\cite{HS05}.

In the hadronic many-body theory (HMBT) approach, starting from a
phenomenological Lagrangian to describe the vacuum properties of the
vector mesons, the in-medium modifications of their spectral properties
are evaluated within finite-temperature/density quantum-field theory,
involving non-perturbative techniques such as the dressing of, e.g., the
pion propagator to account for the modification of the $\rho$-meson's
pion cloud and implementation of interactions of the $\rho$-meson with
mesons and baryons in the medium (for a review, see~\cite{Rapp:1999ej}).
It is characteristic for such models that the various excitations result
in a \emph{substantial broadening of the vector mesons and small mass
  shifts}, i.e., a realization of the ``melting-resonance scenario'' of
CSR. An intriguing property in connection with the model proposed
in~\cite{Rapp:1999us} is that the resulting dilepton-emission rates,
cf. Eq.~(\ref{MT}), match that of the hard-thermal-loop improved pQCD
rate, when both are extrapolated to the expected chiral-phase transition
temperature $T_c \simeq 160$-$190\,\mathrm{MeV}$, i.e., a kind of
``quark-hadron duality''~\cite{Rapp:1999us}. This behavior is consistent
with the smoothness of quark-number susceptibilities in the
corresponding isovector channel across the phase transition in recent
lQCD calculations~\cite{Allton:2005gk}.

Another approach to assess in-medium properties of vector mesons is the
use of empirical scattering amplitudes and dispersion-integral
techniques to assess the in-medium $\rho$-meson propagator via the
$T\varrho$ approximation~\cite{Eletsky:2001bb}.

\section{Dilepton phenomenology in heavy-ion collisions}
\label{sect.phen}

In this Section we compare theoretical models of the in-medium
em. current correlator to experimental results from ultrarelativistic
heavy-ion collisions. For such a comparison, not only detailed models
for the in-medium behavior of the correlation function itself in both
partonic (QGP) and hadronic states of the medium are required, but also
a description of its entire ``thermal evolution'' over which the
rate~(\ref{MT}) has to be integrated to compare to the experimental
observables.

The bulk of hot and dense matter created in ultrarelativistic heavy-ion
collisions at the CERN SPS and RHIC can be successfully described by
ideal hydrodynamics~\cite{Kolb:2003dz}, which implies local thermal
equilibrium. Thus the medium is characterized by a temperature- and
collective-flow field. In \cite{vanHees:2007th,vanHees:2006ng} a simple
thermal model has been used, which after a ``plasma-formation time''
describes the medium by an ideal-gas equation of state of quarks and
gluons which according to recent lQCD
calculations~\cite{Fodor:2004nz,Karsch:2007vw} undergoes a phase
transition at $T_c \simeq 160$-$190 \,\mathrm{GeV}$ to a
hadron-resonance gas. Thermal
models~\cite{Andronic:2005yp,Becattini:2005xt} for the yields of various
hadron species indicate that at a temperature close to the phase
transition of $T_{\mathrm{ch}} \simeq 160$-$175 \, \mathrm{GeV}$,
inelastic reactions within the medium cease, and the corresponding
particle ratios are fixed (\emph{chemical freeze-out}), before the
particles decouple and freely stream to the detector. This \emph{thermal
  freeze-out} occurs at temperatures around $T_{\mathrm{fo}} \simeq
90$-$130 \, \mathrm{GeV}$ (depending on the system size). This evolution
of the medium is implemented through a cylindrical homogeneous fireball
model, including radial flow and longitudinal
expansion~\cite{vanHees:2007th}. The temperature is inferred from the
equation of state (massless gluons and $N_f=2.3$ effective quark flavors
in the QGP and a hadron-resonance gas model in the hadronic phase) and
the assumption of an isentropic expansion in accordance with ideal-fluid
dynamics. Between the pure QGP and hadronic phases a standard volume
partition for a mixed phase is employed. The hadronic phase is
characterized by the build-up of hadron-chemical potentials to keep the
particle-number ratios fixed at the observed values. The largest
uncertainty is the total fireball lifetime which has been adjusted to
the total experimental yield.

\begin{figure}
\begin{center}
\begin{minipage}{0.3\linewidth}
\includegraphics[width=\textwidth]{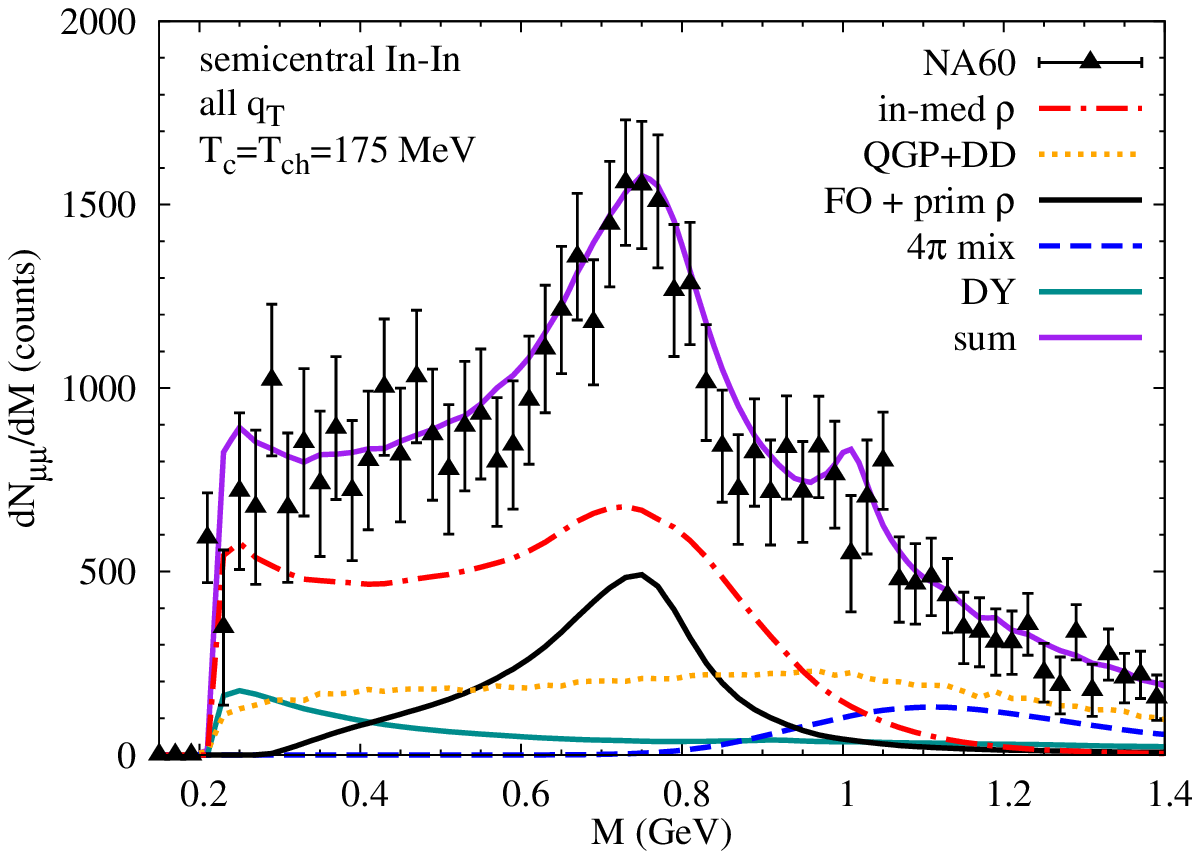}
\end{minipage}\hfill
\begin{minipage}{0.28\linewidth}
\includegraphics[width=\textwidth]{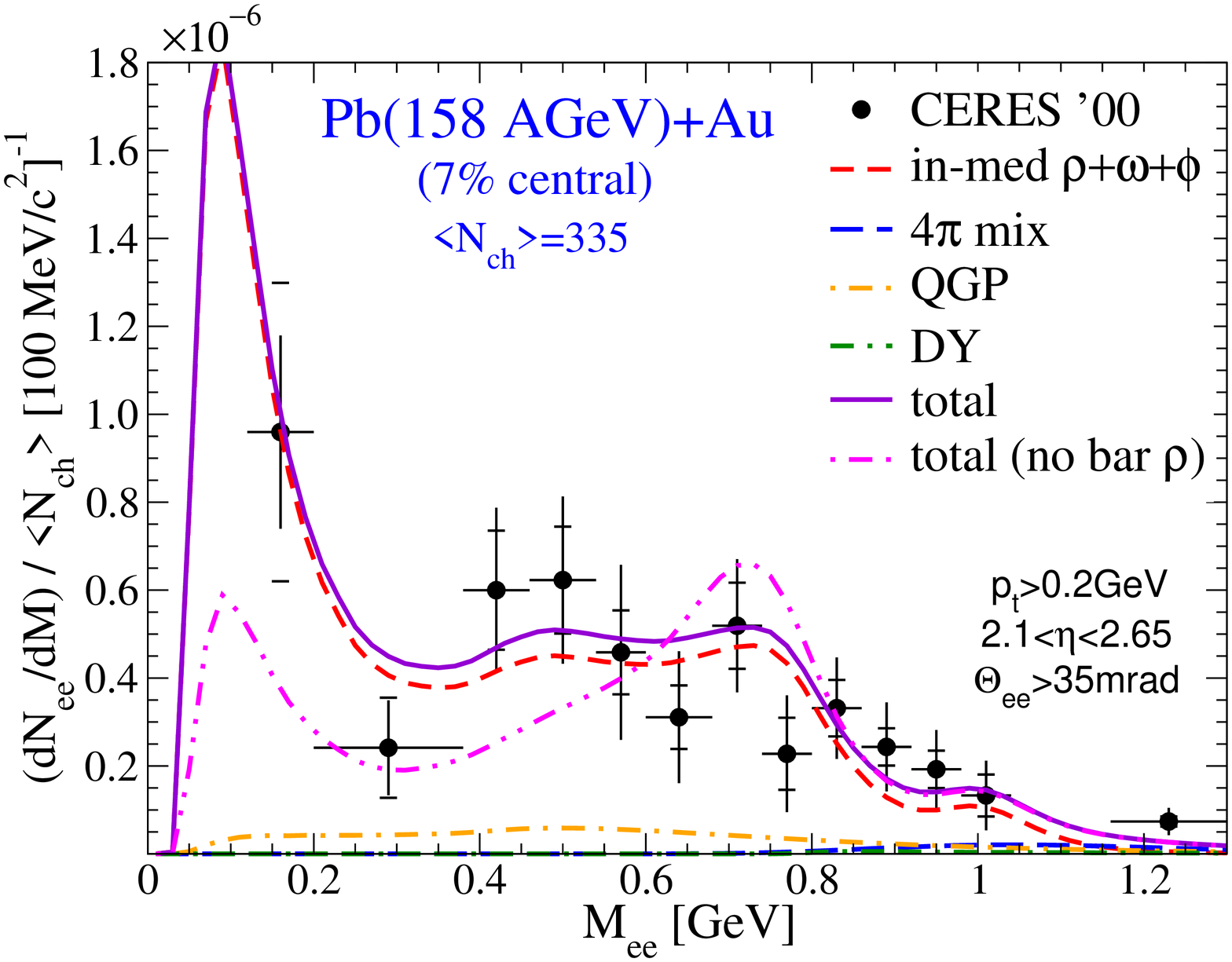}
\end{minipage}\hfill
\begin{minipage}{0.3\linewidth}
\includegraphics[width=\textwidth]{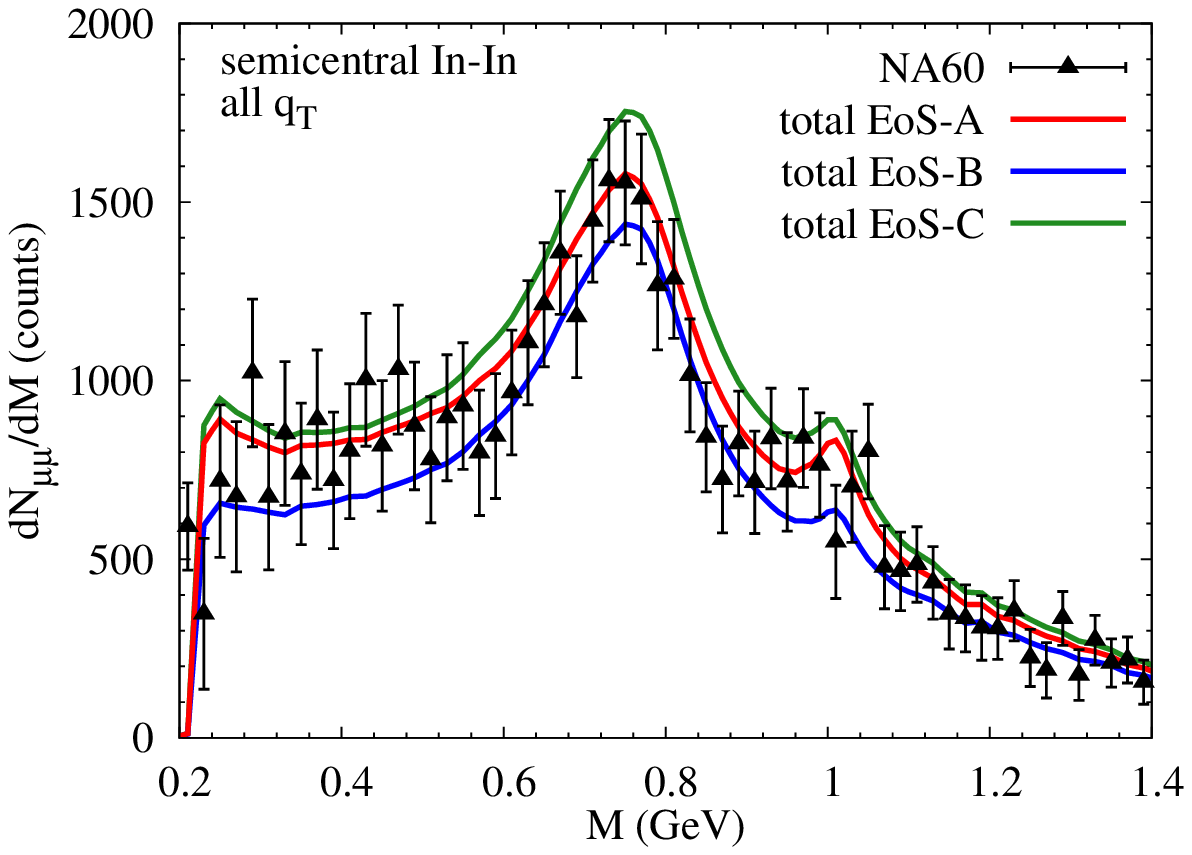}
\end{minipage}
\end{center}
\caption{Left panel: dimuon excess spectrum~\cite{vanHees:2007th} with
  an equation of state with $T_c=T_{\mathrm{ch}}=175\,\mathrm{MeV}$ in
  semicentral $158\,A\mathrm{GeV}$ In-In collisions compared to data by
  the NA60 collaboration~\cite{Arnaldi:2006jq}; middle panel: dielectron
  excess spectrum from the same model for central $158\,A\mathrm{GeV}$
  Pb-Au collisions compared to data by the
  CERES/NA45~\cite{Adamova:2006nu} collaboration. The dash-dotted line
  shows the result with a $\rho$-meson spectral function including only
  medium modifications in a meson gas, underlining the importance of baryon
  effects; right panel: the dilepton excess spectrum based on the
  implementation of different equations of state in the fireball
  evolution (EoS-A: $T_c=T_{\mathrm{ch}}=175\,\mathrm{MeV}$, EoS-B:
  $T_c=T_{\mathrm{ch}}=160\,\mathrm{MeV}$, EoS-C:
  $T_c=160\,\mathrm{MeV}$, $T_{\mathrm{ch}}=160\,\mathrm{MeV}$).}
\label{fig.1}
\end{figure}

\textbf{Invariant-mass spectra.} While earlier dilepton measurements at
the SPS have shown an enhancement of the dilepton yield at invariant
masses in the low-mass region (LMR), $2m_l \leq M \leq 1\,\mathrm{GeV}$,
a definite conclusion concerning the nature of the expected CSR could
not be reached since models for in-medium modifications of the $\rho$
meson based on either the ``dropping-mass'' or the ``melting-resonance''
scenario could describe the data within the experimental mass resolution
and errors. Only recently with the precision reached in the measurement
of dimuon-invariant-mass ($M$) spectra by the NA60
collaboration~\cite{Arnaldi:2006jq} in $158\,\mathrm{GeV}$ In-In
collisions, it could be shown that models predicting a broadening of the
vector mesons with small mass shifts (cf. left panel of
Fig.~\ref{fig.1}) seem to be favored compared to those implementing the
``dropping-mass scenario''. As can be seen in the middle panel of
Fig.~\ref{fig.1} the same model is also consistent with a recent
analysis of $158\,A\mathrm{GeV}$ Pb-Au data on the dielectron-$M$
spectrum by the CERES/NA45~\cite{Adamova:2006nu} collaboration. As shown
by the comparison with the model only implementing mesonic medium
effects, baryonic processes are the prevalent effect leading to the
massive broadening of the $\rho$ meson necessary to explain the observed
dilepton enhancement in the LMR (including the related enhancement below
the two-pion threshold).

While in the LMR the observed access yield over the standard hadronic
cocktail is mostly due to the emission from the medium-modified light
vector mesons, in the intermediate-mass region (IMR),
$1\,\mathrm{GeV}$$\,\leq$$\,M$$\,\leq$$\,1.5\,\mathrm{GeV}$, it is
either dominated by hadronic ``multi-pion processes'' (estimated using
chiral-mixing formulas) or $q\bar{q}$-annihilation in the QGP phase
(given by hard-thermal loop resummed $\bar{q}q$
annihilation)~\cite{vanHees:2006ng,vanHees:2007th}, depending on the
equation of state ($T_c$) as will be described below. As shown in the
right panel of Fig.~\ref{fig.1}, despite small deviations in the overall
yield (which can be adjusted by slight variations of the fireball
lifetime), the spectra are robust against details of the equation of
state within the boundaries of $T_c$ from lQCD calculations and
$T_{\mathrm{ch}}$ from thermal-model analyses. The insensitivity of the
dilepton spectra with respect to the equation of state reflects the
``quark-hadron duality'' of the dilepton rates in the relevant
temperature range close to $T_c$ (see Sect.~\ref{sect.emcc}).
\begin{figure}
\begin{center}
\begin{minipage}{0.21\linewidth}
\includegraphics[width=\textwidth]{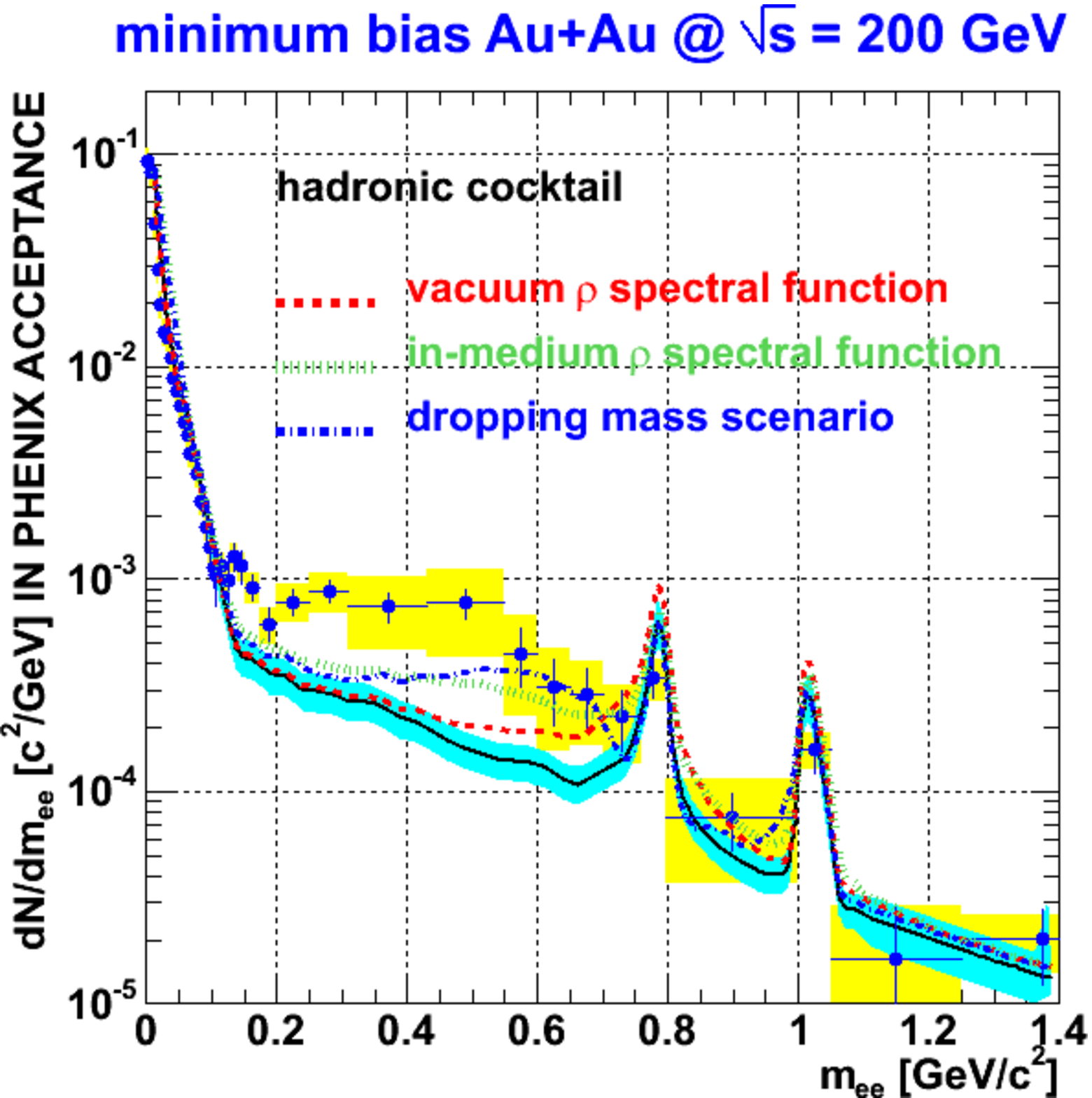}
\end{minipage}\hfill
\begin{minipage}{0.32\linewidth}
\includegraphics[width=\textwidth]{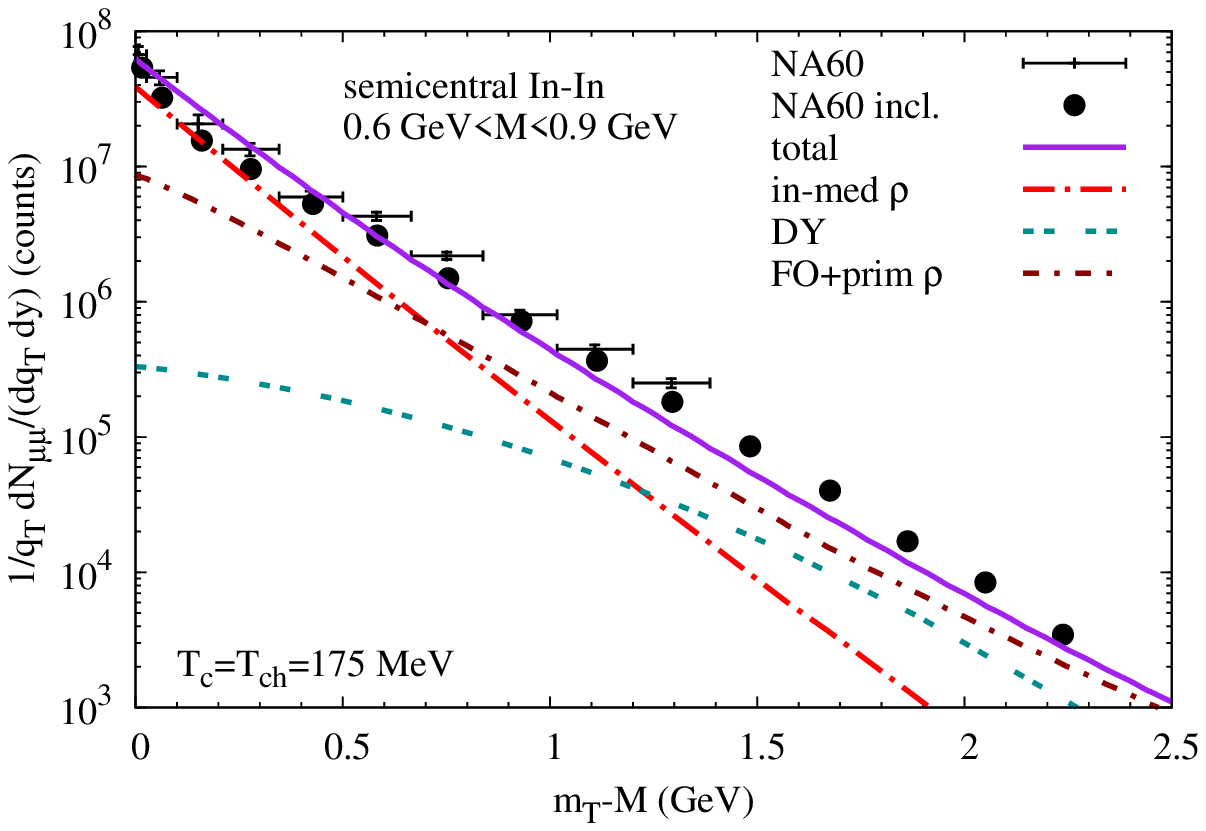}
\end{minipage}\hfill
\begin{minipage}{0.3\linewidth}
\includegraphics[width=\textwidth]{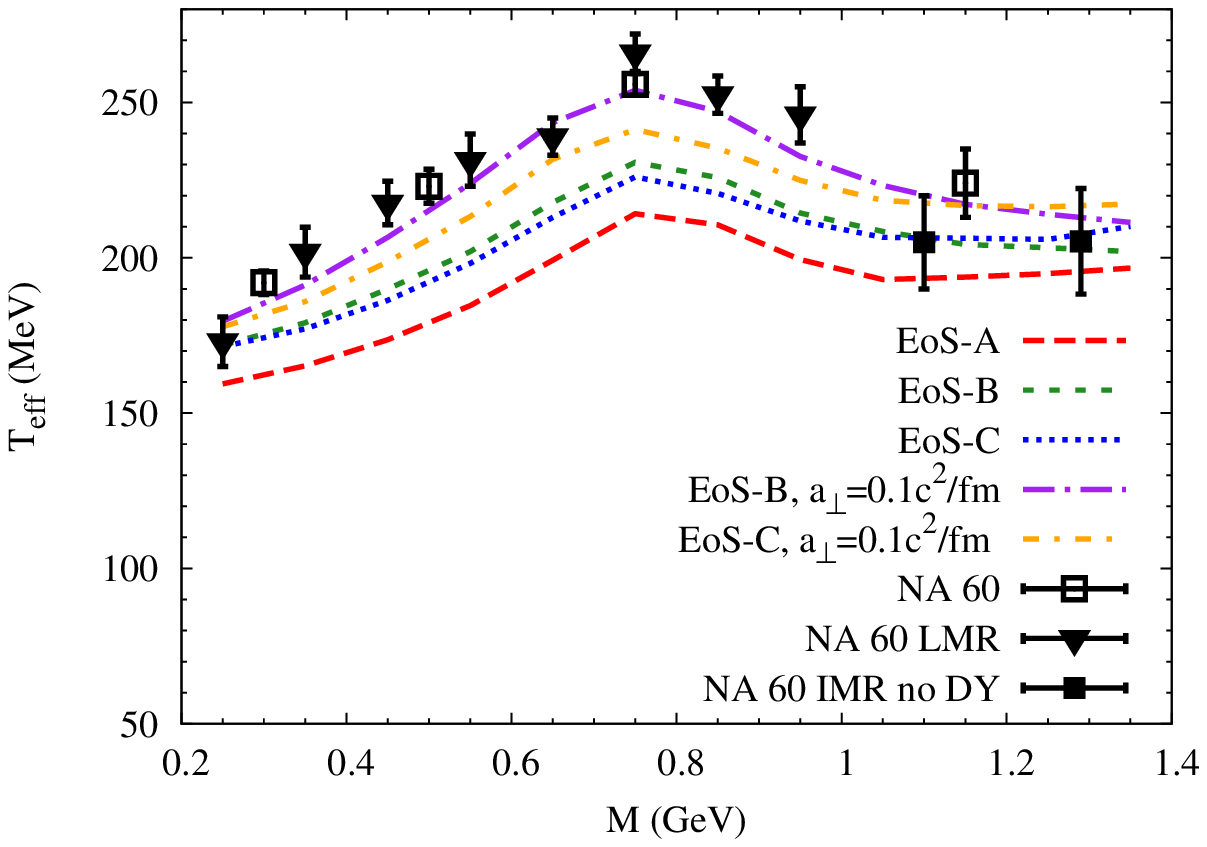}
\end{minipage}
\end{center}
\caption{Left panel: Comparison of dielectron-$M$ spectra based on a
  hadronic-many-body calculation~\cite{Rapp:2002mm} with recent data
  from $200\,A\mathrm{GeV}$-Au+Au collisions at RHIC from the PHENIX
  collaboration~\cite{phenix:2007xw}; middle panel: $m_T$-dimuon
  spectrum in the mass range $0.6\,\mathrm{GeV} \leq M \leq
  0.9\,\mathrm{GeV}$ compared with the NA60 data in $158A\mathrm{GeV}$
  In-In collisions~\cite{Damjanovic:2007qm,Arnaldi:2007ru}; right panel:
  effective slopes from $q_T$ spectra compared to NA60 data for
  different equations of state and transverse acceleration of the
  fireball.}
\label{fig.2}
\end{figure}
Using the spectral functions from the above described chiral-reduction
approach~\cite{Steele:1997tv} within a hydrodynamic description of the
fireball evolution~\cite{Dusling:2006yv,Dusling:2007kh} leads to similar
results for the mass region below and above the $\rho$ region but less
broadening in the resonance region which is to be expected from the
low-density (virial) expansion treatment of the medium effects. The
spectral-functions, based on empirical $\rho$-scattering
data~\cite{Eletsky:2001bb} have been implemented within another
fireball-evolution model (using a cross-over QGP-hadron phase
transition)~\cite{Ruppert:2007cr}, showing results for the $M$ spectra
comparable to those in~\cite{vanHees:2006ng,vanHees:2007th} but with
less enhancement in the mass region below the $\rho$ (particularly below
the two-pion threshold), which may be traced back to the use of the
$T\varrho$ approximation to the medium modifications. In the IMR the
model in \cite{Ruppert:2007cr} shows a large fraction of dilepton
emission from the partonic phase. We close our brief review on
invariant-mass spectra with the remark, that the enhancement of the
dilepton yield in the LMR, observed by the PHENIX collaboration in
$200A\,\mathrm{GeV}$ Au-Au collisions at RHIC~\cite{phenix:2007xw},
cannot be described with the present
models~\cite{Rapp:2002mm,Dusling:2007su}.

\textbf{Transverse-momentum spectra.} The dimuon transverse-momentum
($q_T$) spectra and pertinent effective-slope fits by the NA60
collaboration~\cite{Damjanovic:2007qm,Arnaldi:2007ru} provide
information which is sensitive to the temperature and collective flow of
the medium due to the \emph{blue shift} of the dileptons radiated from a
moving thermal source. While the model in~\cite{vanHees:2006ng}
describes the $q_T$ spectra for $q_T\lesssim 1\,\mathrm{GeV}$ reasonably
well (which is consistent with the agreement in the inclusive $M$
spectra) they were underpredicted at $q_T \gtrsim 1 \, \mathrm{GeV}$
although the fireball parameterization of the temperature and flow agrees
well with results from a hydrodynamic
calculation~\cite{Dusling:2007kh}. Thus sources for dileptons at high
$q_T$ have been investigated, including an improved description of
dileptons from $\rho$ decays after thermal freeze-out which benefit from
the maximal blue shift due to the fully developed transverse
flow~\cite{Rapp:2006cj,vanHees:2007th}. In this connection it is
important to note that the $q_T$ spectrum for emission from a thermal
source cf.~(\ref{MT}) is softer by a Lorentz factor $M/E=1/\gamma$
compared to that from a freely streaming $\rho$ meson due to the
dilation of its lifetime. In former calculations the standard
description for dileptons from freeze-out $\rho$ decays has been to
prolong the fire-ball lifetime for this contribution by $1/\Gamma_{\rho}
\sim 1\,\mathrm{fm}/c$~\cite{Rapp:1999us,vanHees:2006ng}. In addition
decays of hard ``primordial'' $\rho$ mesons, produced in the initial
hard $NN$ collisions which are subject to jet quenching through the
medium but leaving the fireball without equilibrating, have been taken
into account. Another source of hard dileptons is Drell-Yan annihilation
in primordial $NN$ collisions which has been extrapolated to small
invariant masses by imposing constraints from the real-photon
point. Finally, $t$-channel-meson (e.g., $\omega$) exchange
contributions to the yield of thermal dileptons have been
studied. Although the latter show the hardest $q_T$ spectra among all
thermal sources, their absolute magnitude is insufficient to resolve the
discrepancies in comparison to the data at high $q_T$, which however has
improved through the above described more detailed implementation of the
hard (non-thermal) dilepton sources (cf. Fig.~\ref{fig.1} middle
panel). The model of~\cite{Ruppert:2007cr} shows larger slopes (also
compared to the hydrodynamic fireball simulation~\cite{Dusling:2007kh}).

Finally a study of different parameters for the equation of state has
been conducted. The ``standard scenario''
in~\cite{Rapp:1999us,vanHees:2006ng} uses
$T_c=T_{\mathrm{ch}}=175\,\mathrm{GeV}$ (EoS-A). To investigate the
sensitivity of the dilepton spectra with respect to uncertainties in the
equation of state, $T_c$ has been varied within the boundaries of
$160$-$190\,\mathrm{GeV}$ given by different lQCD
calculations~\cite{Fodor:2004nz,Karsch:2007vw} (EoS-B:
$T_c=160\,\mathrm{MeV}$, EoS-C: $T_c=190\,\mathrm{MeV}$), using a
chemical freeze-out temperature of $T_{\mathrm{ch}}=160\,\mathrm{GeV}$
to cover the range of thermal-model fits to hadron-number ratios in
heavy-ion collisions. For EoS-C a chemically equilibrated hadronic phase
between $T_c=190\,\mathrm{MeV}$ and $T_{\mathrm{ch}}=160\,\mathrm{MeV}$
has been assumed. The agreement of the model with the measured $M$
spectra is robust. Variations in the absolute yield can be adjusted by
slight changes in the fireball lifetime. It is important to note that in
the IMR the partition of the dilepton yield in hadronic and partonic
contributions depends sensitively on the equation of state: A scenario
like EoS-B with a low critical temperature results in a long QGP+mixed
phase, leading to a parton-dominated regime in the IMR, while EoS-C with
a high critical temperature describes a hadron-dominated source since
the QGP and mixed phase become shorter. Thus, contrary to suggestions in
the literature~\cite{Arnaldi:2007ru,Specht:2007ez}, a definite
conclusion whether the dilepton yield in the IMR is originating from
partonic or hadronic sources can not be drawn at present. In the right
panel of Fig.~\ref{fig.2} effective slopes extracted from the $q_T$
spectra by a fit to $\dd N_{ll}/(m_T \dd m_T)=C
\exp(-m_T/T_{\mathrm{eff}})$ are shown. The slopes from the calculations
with EoS-B and EoS-C benefit from the larger freeze-out temperature of
$T_{\mathrm{fo}}=136\,\mathrm{MeV}$ compared to
$T_{\mathrm{fo}}=120\,\mathrm{MeV}$ for EoS-A. To reach the measured
effective slopes however, an enhancement of the transverse acceleration
of the fireball (from $a_{\perp}=0.085 c^2/\mathrm{fm}$ to
$a_{\perp}=0.1 c^2/\mathrm{fm}$), leading to larger blue shifts in the
spectra, is necessary. The slopes in the IMR are not so sensitive to the
radial flow since the emission in this region is dominated by radiation
from (either partonic or hadronic, depending on $T_c$) sources at
earlier times where the flow is smaller.

\section{Conclusions and outlook}
\label{sect.concl}

In conclusion, the confrontation of phenomenological models for the
em.~current correlator of strongly interacting matter with precise data
on dilepton emission in high-energy heavy-ion collisions provides a
unique opportunity for a better understanding of the nature of chiral
symmetry restoration. Models based on the application of many-body
theory to phenomenological hadronic models to assess the medium
modifications of the em. current correlator, predicting a strong
broadening of the light-vector-meson spectrum with small mass shifts,
are favored by the data compared to models implementing a dropping-mass
scenario (as, e.g., implied by the intrinsic temperature dependencies of
the model parameters of the generalized hidden-local symmetry model due
to ``Wilsonian matching'' with QCD close to $T_c$~\cite{Harada:2006hu}
although a final confrontation of this particular realization of chiral
symmetry with dilepton data in HICs has to be completed by an
implementation of baryonic interactions). However,
the origin of the large dilepton enhancement in the LMR observed by the
PHENIX collaboration in $200 \,A\mathrm{GeV}$ Au-Au collisions at RHIC
remains unexplained so far.

Future investigations will have to find even closer connections between
em.~observables in heavy-ion collisions and the chiral phase
transition. One possibility is the extension of the hadronic-model
calculations with a detailed study of the in-medium properties of both
the vector and the axial-vector correlator within a chiral framework,
constrained by lQCD calculations of chiral order parameters via
Weinberg-sum rules.

\textbf{Acknowledgment.} I thank the conference organizers for the
invitation to an interesting meeting, R. Rapp for the fruitful
collaboration and S. Damjanovic and H. Specht for stimulating
discussions. This work was supported in part by the U.S. National
Science Foundation under grant no. PHY-0449489.

\begin{flushleft}

\end{flushleft}


\end{document}